\documentclass[a4paper,10pt]{article}
\usepackage[latin1]{inputenc}
\usepackage{amsmath,amssymb}

\begin{document}
\title{
On computations of angular momentum and its flux in numerical relativity
}

\author{
Emanuel Gallo${}^1$\thanks{Member of CONICET; email: egallo@famaf.unc.edu.ar.},
Luis Lehner${}^2$\thanks{
email: lehner@phys.lsu.edu}
and
Osvaldo M. Moreschi${}^1$\thanks{Member of CONICET; email: moreschi@fis.uncor.edu.} \\
\\
${}^1$\small  Instituto de Física (CONICET) \\
\small  FaMAF, Universidad Nacional de C\'{o}rdoba\\
\small Ciudad Universitaria, (5000) C\'{o}rdoba, Argentina. \\
${}^2$\small Department of Physics \& Astronomy, LSU, \\
\small Baton Rouge, LA 70803, USA.
}

\maketitle
\begin{abstract}
 The purpose of this note is to point out ambiguities that appear in the calculation of
 angular momentum and its radiated counterpart when some simple formulae are used to compute
 them. 
We illustrate, in two simple different examples, how incorrect results can be 
obtained with them.
Additionally, we discuss the magnitude of possible errors
in well known situations.
\end{abstract}


\section{Introduction}
It is well known that the notion of total angular momentum in general relativity
is sensitive to the so called problem of supertranslation ambiguities.
This topic involves subtleties and complications inherent in the
infinite dimensional nature of the asymptotic symmetry group of
asymptotically flat spacetimes, namely, the BMS group.
On the other hand, there is a clear need to compute these quantities to
extract valuable physical information in numerical constructed spacetimes to
connect with observations of a variety of sources.
A few proposals exist to do this either
 locally\cite{Komar59,Penrose82,Szabados04,Ashtekar04,Hayward06} 
or asymptotically\cite{Komar59,Bramson75,Prior77,Winicour80,Dray84,Rizzi01,Moreschi04,Szabados08}\footnote{We are not presenting an exhaustive list of references, since this
is not a review paper on angular momentum. We apologize for our arbitrary choice
of references.}.
The dependence of any notion of relativistic angular momentum on
supertranslations has been the matter of numerous studies.
The problem is complex and, unfortunately, most definitions of angular momentum 
either show supertranslation ambiguities or require of additional structure
(at the expense of reducing generality in their application).

In spite of all these difficulties, the need for tackling concrete calculations
of angular momentum in numerically constructed spacetimes has motivated the 
use of simple and practical formulae for the computation of angular momentum. These
formulae lend themselves for a straightforward computation and, in relatively simple scenarios,
have given the expected answers.
One option that has been employed in several works has been suggested in  \cite{Lousto07},
 where it is proposed to compute the angular momentum from the expression
\begin{equation}
  \label{eq:link}
  J_{[i]} = \frac{1}{16 \pi} \Re\text{e} \left\{ \oint_{\Sigma^+} 
   dS \Phi_{[i]} \left[ 2 \Psi^0_1 - 2 \sigma^0 \eth \bar
\sigma^0 - \eth(\sigma^0 \bar \sigma^0)\right] \right\}  ;
\end{equation}
which is a truncated version of the one defined in~\cite{Winicour80}; in fact, a
supermomentum part is missing.
As mentioned above,
supertranslations issues can obscure angular momentum definitions (and expressions derived from them)
and indeed, as we show in the next sections, eq. (\ref{eq:link}) is susceptible to them as
 well.

We discuss in the next examples two simple cases where
expression (\ref{eq:link}) gives an incorrect result.

\section{Examples}

\subsection{A supertranslated and boosted Schwarzschild}\label{subsec:boosted}

Let us consider, for simplicity, the case of a spacetime describing two orbiting
black holes which after merger give rise to a non-spinning black hole.
Furthermore, we consider a case where the final black hole has a kick. This
scenario, can easily be realised with a non-equal mass binary with individual
spins adjusted such that the final spin is zero\footnote{A straightforward calculation
can be employed to adjust the physical parameters so that this
scenario is realized as discussed by the simple model \cite{Buonanno08} or
the fitted expansions presented in \cite{Boyle2007,Rezzolla2007}. The resulting
kick velocity can be estimated by suitable fits as in \cite{Boyle2007,Baker2008,Campanelli2007}.}.
Therefore, the black hole will generically appear in the grid which is supertranslated
from that where the black hole would be at rest.

To simplify the analysis, we further assume that a natural coordinate system
can be constructed such that, asymptotically, the system can be surrounded
by round spheres; that is, one can construct a Bondi type
coordinate system from it. The delicate issue is that this
inertial asymptotic coordinate system results, generically, supertranslated and
boosted with respect to the intrinsic asymptotic coordinate
system which has the spherical symmetry of the remaining black hole.
This in turn, implies that the asymptotic fields acquire a non-trivial
dependence on the supertranslation and boost. To compare what would
be obtained in both systems, we denote with primes the asymptotic coordinate system with
spherical symmetry and unprimed that naturally constructed
in the numerical implementation; 
then one would have the BMS transformation
\begin{align} 
 u &= K(\zeta',\bar\zeta')(u' - \gamma(\zeta',\bar\zeta') )+ 
 \frac{ u_1(u',\zeta',\bar\zeta')}{r'}  \label{eq:tildeu}
+ O\left(\frac{1}{r^2}\right) ,\\
 r &= \frac{r'}{K(\zeta',\bar\zeta')}  \label{eq:tilder}
+ O\left(r^0\right)  ,\\
\zeta & = \frac{a \zeta'+b}{c \zeta'+d} 
+ O\left(\frac{1}{r'}\right) ;\label{eq:tildezeta}
\end{align}
where $K$ is the Lorentz factor associated with the boost,
$\gamma$ determines the supertranslation and $(a,b,c,d)$
are complex constants which determine the boost;
for details see \cite{Moreschi86}.

In what follows it will be convenient to use the notation
\begin{equation}
 e^{2 i \delta}
=
\frac{\bar{ \zeta}_{\bar \zeta'}}{{\zeta}_{\zeta'}}
=
\left( \frac{c \zeta' + d}{\bar c \bar \zeta' + \bar d}\right)^2
.
\end{equation}

Although the calculation of expression (\ref{eq:link}) on a
spherically symmetric section gives the expected null result
we will show next that it gives a non-zero result when calculated
at a supertranslated and boosted section.

As explained elsewhere\cite{Lehner07,Gallo08} the above transformation
implies a transformation of the null frames, and in particular,
the components of the Weyl tensor and the asymptotic shear
are affected. More concretely, one has
\begin{equation}\label{eq:forma1psi1}
\begin{split}
 \Psi_1^0 
=
\frac{e^{i \delta}}{K^3}
\left( \Psi_1^{'0} 
-3 \frac{\eth'\, u}{K} \Psi_2^{'0}
+3 \left(\frac{\eth'\, u}{K}\right)^2 \Psi_3^{'0}  
+ \left(\frac{\eth'\, u}{K}\right)^3 \Psi_4^{'0}  
\right) 
.
\end{split}
\end{equation}
Only $\Psi_2^{'0}=-m$ is different from zero in the spherically symmetric 
frame; therefore one has
\begin{equation}\label{eq:psi1}
 \Psi_1^0 =
\frac{3m}{K^4}
e^{i \delta}
\;
\eth'\, u
;
\end{equation}
where $m$ is the mass of the resulting black hole.
Similarly the shear is
\begin{equation}\label{eq:shear}
 \sigma^0 = 
-\frac{1}{K}
e^{i 2 \delta}
\eth^{'2} \gamma
.
\end{equation}

Let us now calculate expression (\ref{eq:link})
 in detail, at a $u=$const. section, and discuss what is obtained.
First, note that
the factors $\Phi_{[i]}$ in the integrand, are spin weighted quantities
with spin weight -1. Therefore they can be expressed in terms of an
edth bar operator acting on a spin weight 0 quantity, namely
 $\Phi_{[i]}=\bar\eth a_{[i]}$ ;
where, to pick the angular momentum components $a_{[i]}$ should be imaginary\cite{Moreschi04}. 
Consequently, each  $a_{[i]}$ is a quantity of spin weight 0 satisfying
$
 \eth \bar\eth a_{[i]} = -a_{[i]}
.
$

We should now express the integration appearing in (\ref{eq:link})
in terms of the primed, spherically symmetric Bondi system.
Let us observe that for all these quantities $a_{[i]}$, 
due to the transformation properties of the edth operator\cite{Newman66},
one has
\begin{equation}
 \bar\eth a = \frac{e^{- i \delta}}{K} \bar\eth' a' ;
\end{equation}
where $a'= a'(\zeta', \bar\zeta')$ is understood in terms of the
primed angular coordinates. Next, one must also take into account that
$dS = K^2 dS'$, then the first term of (\ref{eq:link}), involving $\Psi_1^0$,
is
\begin{equation}
  \label{eq:linka}
   \Re\text{e} \left\{ \oint_{\Sigma^+} 
    \Phi_{[i]}  \Psi^0_1  dS\right\} 
=
   \Re\text{e} \left\{ \oint_{\Sigma^+} 
\frac{e^{- i \delta}}{K} \Phi'_{[i]}  
\frac{3m}{K^4}
e^{i \delta} \; \eth'\, [K(u'- \gamma)]
K^2 dS'
\right\} 
.
\end{equation}
In order to simplify the discussion we can assume that the calculation
is performed at the section $u=0$, which is equivalent to
$u'= \gamma$;
therefore, one has
$ \eth'\, [K(u'- \gamma)]
=
(u'- \gamma)\eth'\, K
-
K
\eth'\, \gamma
=
-
K
\eth'\, \gamma
$;
which implies
\begin{equation}
  \label{eq:linka1}
   \Re\text{e} \left\{ \oint_{\Sigma^+} 
    \Phi_{[i]}  \Psi^0_1  dS\right\} 
=
   -\Re\text{e} \left\{ 3 m \oint_{\Sigma^+} 
\Phi'_{[i]}  
K^{-2}
\; \eth'\, \gamma
 dS'
\right\} 
.
\end{equation}
Since the factor $K^{-2} \eth' \gamma$ has spin weight 1,
one again can think of a spin-weight 0 potential $A$ such that
 $\eth'A = K^{-2} \eth' \gamma
$.
Therefore, if the quantity $A$ were real, this term would not have
any contribution, since one would have to evaluate
\begin{equation}
\begin{split}
  \label{eq:linka2}
   \Re\text{e} \left\{ \oint_{\Sigma^+} 
    \Phi_{[i]}  \Psi^0_1  dS\right\} 
=&
   -\Re\text{e} \left\{ 3 m \oint_{\Sigma^+} 
\bar\eth' a'_{[i]}
\eth' A
 dS'
\right\} 
=
   \Re\text{e} \left\{ 3 m \oint_{\Sigma^+} 
\eth' \bar\eth' a'_{[i]}
 A
 dS'
\right\} \\
=&
  -\Re\text{e} \left\{ 3 m \oint_{\Sigma^+} 
 a'_{[i]}
 A
 dS'
\right\} 
=0
;
\end{split}
\end{equation}
since $a'_{[i]}$'s are imaginary and we are assuming at the moment that $A$ is real.
Unfortunately, it is not difficult to prove that actually
$A$ has an imaginary part; and furthermore, this imaginary
part is only zero when the gradients of $\frac{1}{K^2}$ and
$\gamma$ are proportional; something that is not true in
our case. 

To see this, we next study the equation $\eth A =v^2\eth \gamma$;
where $v=K^{-1}$,
 and
assuming $A$ is a real quantity we will reach a contradiction.
Notice that for a real spin weight zero field $A$,
 $\text{Im}\left[\bar\eth\eth A\right]=0$; since
$\bar\eth \eth$ is a real operator when acting on a spin weight zero quantity.
This last relation can be written in terms of $v$ and $\gamma$ as 
 \begin{equation}\label{eq:relacion apendice}
  \bar\eth v^2\eth \gamma-\bar\eth \gamma\eth v^2=0.
 \end{equation}
Let us denote the gradients on the sphere of the quantities $v^2$ and $\gamma$ by
$\nabla_a v^2$ and $\nabla_b \gamma$. Then the `cross product' of these vectorial
quantities is defined by $\epsilon^{ab}\nabla_a v^2 \nabla_b \gamma$;
where $\epsilon^{ab}$ is the surface element of the unit sphere.
At this point it is convenient to express the gradients in terms of the
edth operator; for example for $v^2$ one has
 \begin{equation}
 \nabla_a v^2 =-(\eth v^2)\bar m_a-(\bar\eth v^2 ) m_a ;
 \end{equation}
and similarly for $\gamma$. Then the cross product is
\begin{equation}
 \epsilon^{ab}\nabla_a v^2 \nabla_b \gamma =
\left( \bar\eth v^2\eth \gamma-\bar\eth \gamma\eth v^2\right) 
\epsilon^{ab} m_a \bar m_b ;
\end{equation}
which should be zero if $A$ were real. But this is true only if the gradients
are proportional.
Consequently the term (\ref{eq:linka1}) has a 
non-zero contribution to
the quantity $J_{[i]}$.\\

We now turn our attention to the second term in expression (\ref{eq:link})
which involves quadratic terms in $\sigma$. 
One could also in this case think of a potential $X$ such
that $\sigma^0 = \eth'^2 X$. Then, one can prove that
if $X$ were real, then the second term of (\ref{eq:link})
involving the $\sigma's$ would not contribute to the
integral\cite{Bramson76}. However, following a similar line of 
arguments as above, one can prove that $X$ will have an imaginary part if
\begin{equation}
-\frac{1}{K}e^{i 2 \delta} \neq 1 \, ,
\end{equation}
--recall this is the factor appearing in (\ref{eq:shear})--.

Therefore, this second term will also have a non-zero contribution
to the quantity $J_{[i]}$; and since it includes independent information
not contained in the first term
it can not cancel the first contribution.

Let us discuss further what we have found up to this point. If
the adopted coordinate system, for instance the one most
naturally constructed from a numerical simulation, results
supertranslated and boosted with respect to the
final frame adapted to the spherically symmetric final black hole
an incorrect result will be obtained for the angular momentum of
the system. Notice that this conclusion is more subtle than the obvious
`Newtonian'-type observation stating that the angular momentum
calculated at frames translated with respect to each other 
are different. In fact, even if the coordinate system is only boosted and
supertranslated --{\bf\em without a translation}--, with respect 
to the one adapted to the spherically symmetric final black hole
then $J_{[i]}$ turns out to be different from zero in the Schwarzschild geometry.
This is to be contrasted with what would be the case within 
special relativity paradigm where the absence of a translation 
does not affect the angular momentum calculation since 
the Lorentz transformation of a zero tensor is a zero tensor.

\subsection{A radiating second order solution}
As a second example let us consider the spacetime given by the metric
\begin{equation} \label{eq:rt-o2}
 ds^{2} =
\left( -2\frac{\dot V}{V}r
+K_V-2\frac{M}{r} \right) du^{2}
+2\;du\;dr-\frac{r^{2} }{V^2 P_0^{2} } d\zeta \;d\bar{\zeta }
;
\end{equation}
where a dot denotes a derivative with respect to $u$,
$K_V$ is the Gaussian curvature,
 given by
\begin{equation}
\begin{split}\label{eq:kurv1}
K_{V}&= 2 V ~\overline{\eth }\eth\, V - 2 \eth V~%
\bar{\eth} V + V^{2} 
,
\end{split}
\end{equation}
of the 2-metric
\begin{equation}\label{eq:desphere}
dS^{2}=\frac{1}{V^2 \, P_0^{2}}\;d\zeta \;d\bar{\zeta} ;
\end{equation}
and $P_0(\zeta,\bar{\zeta})$ is the conformal factor of the unit sphere.
This metric was used by Robinson and Trautman to characterize spacetimes
satisfying Einstein equations admitting a congruence of null geodesic
without shear and twist\cite{Robinson62}.

The scalar $V(u,\zeta,\bar{\zeta})$ can be expressed in the form
$V=1+\Delta(u,\zeta,\bar{\zeta})$; so that $\Delta =0$ would
correspond to the Schwarzschild spacetime.
The function $\Delta$ is required to satisfy the equation
 \begin{equation} \label{eq:rteq}
-3\;M\;\dot{\Delta} =
(1+4 \Delta) \;\eth^{2}\bar \eth^2 \Delta
-\eth^{2} \Delta  \bar\eth^2 \Delta .
\end{equation}

This spacetime is known as a Robinson-Trautman geometry\cite{Dain96}.
By requiring (\ref{eq:rteq}) one obtains a solution of vacuum
Einstein equations in second order in $\Delta$.
The Robinson-Trautman spacetimes are understood as representing
a central object which has been perturbed and are known to
decay in time to the Schwarzschild geometry.

Suppose one has obtained this spacetime and would like to calculate
the quantity expressed in  (\ref{eq:link}) at the section $u=u_0$. 
Then, since the asymptotic
coordinate system $(u,\zeta,\bar\zeta)$ is not inertial (not Bondi);
one should look for the transformation to a Bondi frame.
Let us choose a Bondi coordinate system 
$(\tilde u,\tilde \zeta,\bar{\tilde\zeta})$
such that $\tilde u=0$ coincide with the section $u=u_0$.
With this choice one has that on this particular section
the Bondi shear is zero; $\tilde \sigma^0=0$\cite{Gallo08}.
Furthermore, since on this section the null frames
are proportional, one also has $\tilde\Psi_1^0 =0$.
Therefore in this section one obtains $J_{[i]} = 0$.

However in any other Bondi section given by $\tilde u \neq 0$
one will obtain $J_{[i]} \neq 0$.
This can be seen due to two reasons. First one can notice
that the shear with respect to the Bondi system is not stationary;
in fact one can prove that 
$\tilde\sigma_{,\tilde u}
= (\eth^2 V){V^{-1}}$\cite{Dain96}.
One can convince oneself that $\tilde\sigma$ can not be expressed in
terms of a real potential $X$, that is, $\tilde\sigma=\eth^2 X$
as studied in the previous section. In other words such an $X$
has an imaginary term.
Therefore one deduces that there is always
a flux, which in general will include the time variation of $J_{[i]}$,
depending on the arbitrary initial data $V$.
On the other hand one can also infer that $\Psi_1^0$ will
also be different from zero, at any other Bondi section;
due again to the existence of radiation.

Notice that $J_{[i]} \neq 0$ at any other Bondi section $\tilde u \neq 0$
can also be deduced by taking (Bondi) time derivatives of $J_{[i]}$.
One can see that at $\tilde u = 0$, one has ${J_{[i]}}_{,\tilde u}=0$,
${J_{[i]}}_{,\tilde u \tilde u}=0$  but
${J_{[i]}}_{,\tilde u \tilde u \tilde u}\neq 0$.

The weaknesses of formula (1) to calculate the angular momentum 
also give rise to odd results. Notice that the
asymptotic value of $J_{[i]}$, for $\tilde u\rightarrow \infty$ is
zero;  since in this regime, $\tilde u$ will be
supertranslated with respect to the $(u,\zeta,\bar\zeta)$ frame;
but not boosted (see the previous section).
So, in this case one would have the curious situation that the initial 
value of $J_{[i]}$ is zero, a non-zero flux to the future 
of the initial section and an ending with a zero value for $J_{[i]}$
in the asymptotic regime.
This alone makes it difficult to identify $J_{[i]}$ with a
physical notion of intrinsic angular momentum.

One instead could use the Bondi system $(u^*,\zeta^*,\bar{\zeta^*})$
which is defined so as to coincide asymptotically with the R-T
system $(u,\zeta,\bar\zeta)$ in the regime $u\rightarrow \infty$. Then one could calculate
the quantity $J_{[i]}$ for any $u^*=$constant. In this
case one would obtain that in the limit $u^* \rightarrow \infty$,
$J_{[i]}$ would vanish. However for any finite
$u^*=$const., the evaluation of $J_{[i]}$ would give a non zero value.

This again stresses the difficulties that one encounters in trying to use
the expression (\ref{eq:link}) to obtain intrinsic information of
the central object.

\section{Estimate of the error in spin calculations}
In order to provide a different perspective to the quantity $J_{[i]}$,
in the next subsection we recall an unambiguous definition of 
total intrinsic angular momentum at future null infinity.
Actually since this is the only definition we know of that 
solves simultaneously the supertranslations ambiguities for
center of mass and 
intrinsic angular momentum\cite{Moreschi04} --without assuming
further structure--; we will use it as
reference for comparison.

\subsection{An unambiguous definition of intrinsic angular momentum}
Arguably one of the most valuable physical information aimed to be extracted 
from a given spacetime
is that of {\em intrinsic angular momentum} as it relates directly with observable signatures
at both gravitational and electromagnetical signals produced in a number of systems.
Tied to defining such quantity is the
problem of defining the  center of mass of the system with respect to which the intrinsic
angular momentum can be defined; consequently, the definition of both concepts must come together. 
A solution of these difficulties has been formulated in~\cite{Moreschi04}; where the expression
to calculate the charges associated to BMS generators is given by,
\begin{equation}
  \label{eq:charge}
\begin{split}
Q_{S_{cm}}(w)=&  
\; \Re\text{e} \left\{8 \int_{S_{cm}}\Bigl( - \, w_{2} \left( \Psi_{1}^{0} 
+ 2\sigma_0 \eth\bar{\sigma_0 } 
 +  \eth\left( \sigma_0 \bar{\sigma_0 } \right)
\right) 
 +\, 2 \,w_{1} \left( \Psi _{2}^{0} +\sigma_0 \dot{\bar{\sigma_0 } } +
\eth^{2} \bar{\sigma_0 } \right) \Bigr) \;dS^{2}\right\}
\end{split}
\, ;
\end{equation}
where  $w_{1}$ and $w_{2}$ are 
components of the two form $w$ that is determined by the particular
generator of BMS transformations. As with the expression appearing
in ~\cite{Winicour80}, one can use this type of expression to 
calculate the total momentum, total supermomentum and
total angular momentum. However, the notion of 
{\bf\em intrinsic angular momentum} can only be obtained if
one calculates the angular momentum on the 
{\bf\em center of mass sections}, here denoted by $S_{cm}$.

If one calculates the angular momentum in a radiating spacetime
at any other section, one would obtain a quantity with
angular momentum reminiscence but with unclear intrinsic physical
meaning.

\subsection{How important are in practice these considerations in the computation of angular momentum?}

We want to estimate now the errors in the value of angular momentum by using 
an unfortunate choice of frame and section.
To do so, we want to have an estimate of how supertranslated are the ``center of mass''
sections  with respect to the sections that are adapted to the coordinates used in
numerical computations. 

The astrophysical model that we have in mind is similar to the system that was presented
in subsection \ref{subsec:boosted}, i.e, we have some compact objects (for example a binary system) that are
undergoing a merge, and where the coordinate grid is in some way following the system
until the moment of merger. After such event, 
in general the system will end up with a unique boosted (quasi)-stationary compact object
(in the case of black hole collision, we will finish with something similar to a boosted
Schwarzschild or Kerr black hole).
Then, the new ``center of mass'' sections will be supertranslated and boosted with respect
to the originals (those adapted to the numerical grid).

Sections that characterise rest frames are known as `nice' sections\cite{Moreschi88}.
The prescription to find rest frame sections uses a particular notion
of supermomentum.

The supermomentum at S is defined\cite{Moreschi88} by
\begin{equation} \label{rf1}
P_{lm} (S)\equiv -\frac{1}{\sqrt{4\pi } } \int\limits_{S}Y_{lm} (\zeta
,\bar{\zeta } )\;\Psi\;dS^{2};
\end{equation}
where
$
\Psi= \Psi_2 +\sigma \dot{\bar{\sigma } } +\eth^2\bar{\sigma }
$
and $Y_{lm}$ are spherical harmonics.

Given another section  $\tilde{S}$ of future null infinity,
one can find another Bondi system $(\tilde{u} ,\tilde{\zeta } ,\bar{\tilde{\zeta } } )$
 such that 
 $\tilde{S} $
 is determined by 
 $\tilde{u} =0$. 
The relation between the new and the original Bondi system is 
given by a BMS transformation, as given by
(\ref{eq:tildeu}), (\ref{eq:tilder}) and (\ref{eq:tildezeta});
where now we use tildes for the new coordinates and un-tilde
for the original coordinates.
Note that 
 $\tilde{S} $
 can also 
be determined by 
 $u=\gamma (\zeta ,\bar{\zeta } )$
.

The section 
 $\tilde{S} $
 is said to be of type nice\cite{Moreschi88} if all the `spacelike' components of the supermomentum, 
when calculated with respect to the adapted Bondi system, are 
zero;
\begin{equation} \label{rf5}
\tilde{P}_{lm} (\tilde{S} )=0  \quad  \text{for} \quad       l\neq 0;
\end{equation}
 and therefore the only non-vanishing one, 
 $\tilde{P} _{00} (\tilde{S} )$, coincides with the 
total Bondi mass at $\tilde{S}$.

One can prove that  
 $\Psi $
 transforms under a BMS transformation 
as
\begin{equation} \label{rf9}
\tilde{\Psi } =\frac{1}{K^{3} } \left( \Psi -\eth^{2}\bar \eth^{2} \gamma \right).
\end{equation}
Then, for a section to correspond to a `nice' section it must 
obey $\tilde{\Psi }=\text{constant}$ as otherwise some moments of $P_{lm}$,
for $l\neq 0$, will be non-zero. 
The nice section equation can be understood 
as a condition for  $\gamma (\zeta ,\bar{\zeta } )$ and $K$.

 It was indicated 
in ref. \cite{Moreschi88} that equation (\ref{rf9}) can be expressed by
\begin{equation} \label{eq:formaestimacion}
\eth^2 \bar \eth^2 \gamma =\int^{\gamma}_0\dot\Psi(u',\zeta,\zeta')du'+
\Psi(u=0 ,\zeta ,\bar{\zeta } )+K(\gamma ;\zeta ,\bar{\zeta } )^{3}M(\gamma )
.
\end{equation}
where $M(\gamma)$ is the total mass at the section $u=\gamma$ and  $K(\gamma ;\zeta ,\bar{\zeta } )$ 
is the conformal factor of 
the BMS transformation that aligns its timelike generator
(defined as
 $\tilde{p} _{00} = Y_{00} \frac{\partial}{\partial \tilde u}$\cite{Moreschi88})
with the Bondi momentum at 
 $u=\gamma (\zeta ,\bar{\zeta } )$.

Note that $\dot\Psi=\dot\sigma^0\dot{\bar\sigma}^0$, and 
then it is proportional to the content of gravitational radiation.
Now, let us introduce a parameter $\lambda$ to measure this content of 
radiation, i.e., 
let the integrated flux, appearing in the first term of the right hand side
of (\ref{eq:formaestimacion}) be of order $\lambda$.

Suppose then, that we start by describing our compact bodies system 
with sections that are close to `nice' sections. Then, after the merger, 
a new compact object is obtained, and due to the emission of gravitational
 radiation, the new `nice' sections that `follow' the system will, in general, not only be
boosted but also supertranslated by a quantity $\gamma$.
Now, let $v$ be the kick velocity of the final compact object; them, 
one can observe that $K^3\approx 1+O\left(v\right).$ 
On the other hand, by computing the flux of the Bondi momentum expression 
we can see that
\begin{equation}
 v\approx O\left(\frac{\lambda}{M}\right);
\end{equation}
 and $\Psi(u=0 ,\zeta ,\bar{\zeta } )\approx-M,$
then 
\begin{equation}
 \eth^2\bar\eth^2\gamma \approx O(\lambda)+O(vM)\approx O(vM).
\end{equation}

From these estimates, and recalling that $\Psi^0_2\approx O(M)$, 
and using eq.(\ref{eq:psi1}), we have that $\Delta\Psi^0_1 \approx O(vM^2)$.

This will be the order of the error that one would have in the computation of angular momentum by
not adopting the correct sections. We conclude then, that the errors will 
be, at least, as large as the relativistic velocities of the astrophysical 
system, i.e., $ \Delta J_{[i]}\approx O(vM^2)$.
Consequently one can understand why the use of the flux of eq.~(\ref{eq:link}),
as calculated in \cite{Lousto07}, employed in several works
\cite{Baker:2008mj,Campanelli:2008nk,Lousto:2008dn,Dain:2008ck,Lousto:2007db} (where the velocities of the astrophysical systems under consideration are low, and therefore the error estimate $\Delta J_{[i]}$ is small)
can produce sensible results.
However one should be cautious with the use of formulas like (\ref{eq:link}) in cases where
the individual objects move relativistically (as in the studies of high speed
collisions~\cite{Sperhake08}), 
large production of gravitational radiation or producing an object with a small final spin 
(where the errors described above will be important).

Finally, we stress that all these considerations are true if one assumes that the numerical
grid and gauge gives rise to an extraction frame/coordinate system which is of Bondi type 
(which in general will not be the case). If it is not of
this type, one must  consider extra gauge effects like those reported in \cite{Lehner07}
and \cite{Gallo08}  in the computation of radiation flux and total momentum respectively.

\section{Final comments}
The difficulties shown above are related to the nature of the notion of
angular momentum.
In special relativity the angular momentum is expressed in terms
of an antisymmetric tensor
\begin{equation}
 J^{ab} = S^{ab} + P^a R^b - R^a P^b ;
\end{equation}
where $P^a$ is the total momentum, $S^{ab}$ is orthogonal to the momentum
and $R^a$ shows the dependence of the angular momentum with respect
to some reference point, or origin.

The tensor $S^{ab}$ does not depend on the choice of origin and therefore
is the intrinsic angular momentum. At the center of mass line one has
$J^{ab} = S^{ab}$ and the original expression of the angular momentum
gives the intrinsic value.

The expression (\ref{eq:link}) does not have this translation/supertranslation
dependence; and therefore its usage is ambiguous since there
is no prescription of what a center of mass is.
Instead, we have shown that one could use a definition of 
angular momentum\cite{Moreschi04} that does show this translation/supertranslation
dependence; and furthermore, it can be used to define center of mass sections,
in which the calculation of angular momentum gives 
an unambiguous intrinsic angular momentum.
In particular the application of this intrinsic angular momentum
to the two examples presented above, gives
zero at any center of mass section.

Incidentally, note also that for general spacetimes, 
inferring how much angular momentum is radiated by differentiating 
with respect to time eq. (\ref{eq:link}) will give results strongly
affected by the issues discussed here.



\end{document}